\documentclass[10pt,conference,final]{IEEEtran}

\usepackage[noadjust]{cite}

\usepackage{graphicx} 
\usepackage{physics}
\usepackage{amsmath}
\usepackage{amssymb}
\usepackage[hidelinks,colorlinks=true,linkcolor=blue,citecolor=blue]{hyperref}

\usepackage{tikz}
\usetikzlibrary{quantikz}
\usepackage{adjustbox}

\usepackage{balance}
\usepackage{caption}
\usepackage{subcaption}

\begin{document}
\title{\Large \bf  
Fast Quantum Process Tomography via Riemannian Gradient Descent
\thanks{This work was partially supported by the NSF grant CCF-1908131.}
}

\author{
    \IEEEauthorblockN{Daniel Volya, Andrey Nikitin and Prabhat Mishra }
\IEEEauthorblockA{University of Florida, Gainesville, Florida, USA}
}
\maketitle

\begin{abstract}
Constrained optimization plays a crucial role in the fields of quantum physics and quantum information science and becomes especially challenging for high-dimensional complex structure problems.
One specific issue is that of quantum process tomography, in which the goal is to retrieve the underlying quantum process based on a given set of measurement data.
In this paper, we introduce a modified version of stochastic gradient descent on a Riemannian manifold that integrates recent advancements in numerical methods for Riemannian optimization.
This approach inherently supports the physically driven constraints of a quantum process, takes advantage of state-of-the-art large-scale stochastic objective optimization, and has superior performance to traditional approaches such as maximum likelihood estimation and projected least squares.
The data-driven approach enables accurate, order-of-magnitude faster results, and works with incomplete data.
We demonstrate our approach on simulations of quantum processes and in hardware by characterizing an engineered process on quantum computers. 
\end{abstract}
\begin{IEEEkeywords}
Quantum process tomography, quantum characterization, quantum computing, Riemannian optimization. 
\end{IEEEkeywords}

\section{Introduction}

Quantum computers present a theoretical exponential speedup over classical computers for various problems. However, the capability of these quantum computers is largely limited by environmental noise. Numerous quantum devices experience undesired interactions with internal and external degrees-of-freedom, resulting in decoherence \cite{aruteQuantumSupremacyUsing2019, madsenQuantumComputationalAdvantage2022}.
Although there are promising algorithmic approaches to preserve coherence using quantum error correcting codes, their effectiveness further depends on appropriate understanding and formulation of the underlying noise \cite{fowlerSurfaceCodesPractical2012, guillaudRepetitionCatQubits2019}.
Thus, accurately and effectively describing the noise in current quantum computing hardware continues to be a crucial component for achieving their long-term technological potential.

\begin{figure}[t!]
    \centering
    \begin{subfigure}[b]{\linewidth} 
         \centering
        \includegraphics[width=0.9\textwidth]{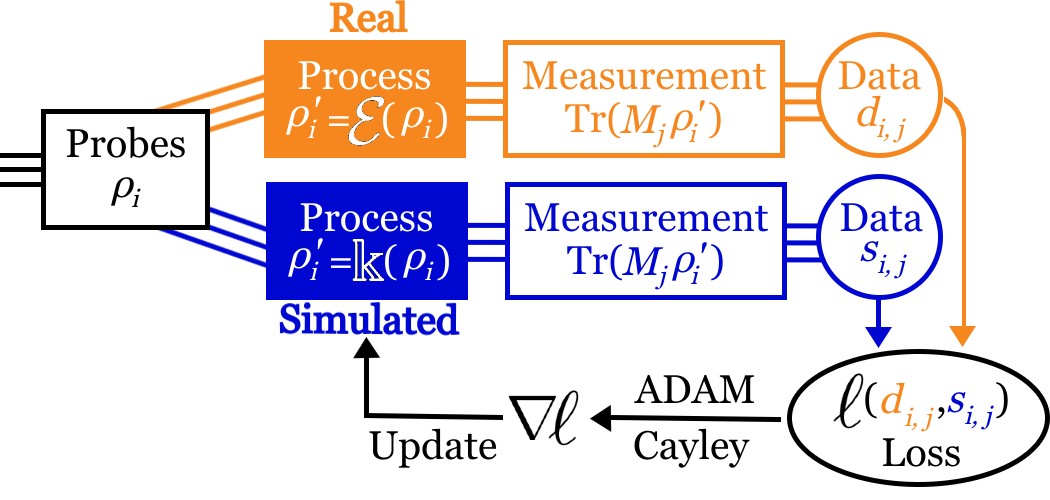}
        \vspace{-0.1in}
        \caption{Quantum process tomography estimates the blackbox process via measurement data $d_{i.j}$. The data is the result of conducting measurements $M_j$ on states produced by a channel $\mathcal{E}$ acting on known probe states $\rho_i$.}
        \label{fig:gradDescent}
     \end{subfigure}
     \hfill
     \begin{subfigure}[b]{\linewidth}
         \centering
         \includegraphics[width=0.9\textwidth]{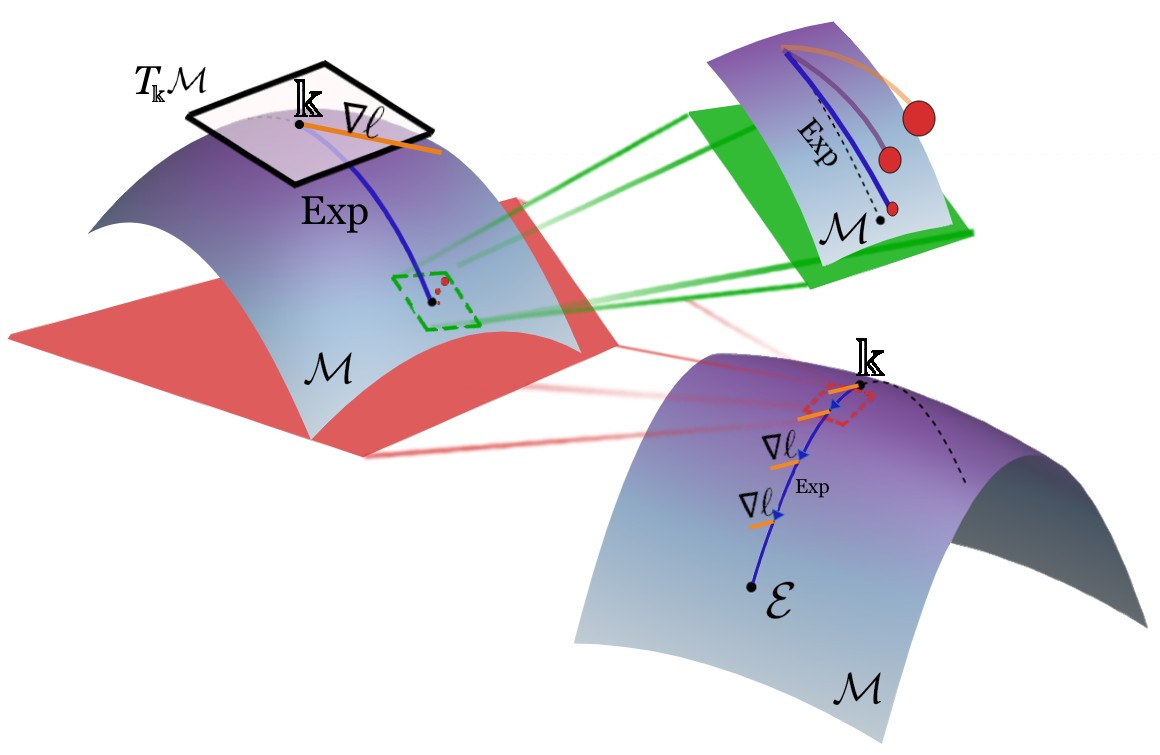}
         \vspace{-0.05in}
         \caption{We estimate a quantum process via gradient descent of an objective loss function $\ell(\cdot)$ on a Riemannian manifold, defined as a set of Kraus operators $\Bbbk$. We adapt an iterative Cayley transformation to perform a retraction of the gradient $\nabla\ell$, approximating the exponential map of a tangent vector onto the manifold.  }
         \label{fig:manifold}
     \end{subfigure}
     \caption{Overview of proposed quantum process tomography.}
\vspace{-0.2in}
\end{figure}

Certain noise can be characterized as a quantum channel, which is a completely positive and trace-preserving mapping between quantum states. Quantum channels are often represented as a set of matrices that transform a quantum state $\rho$ to another state $\rho^\prime$. As shown in Fig.~\ref{fig:gradDescent}, given the initial and resulting states, it is possible to reconstruct the quantum channel through tomography.
Tomography refers to the procedure of creating a model by merging various partial cross-sections or slices \cite{chuangPrescriptionExperimentalDetermination1997}. Although each slice provides a limited viewpoint, the data obtained from numerous independent experiments allows for the construction of a comprehensive model for the system.
This paper specifically focuses on quantum process tomography (QPT), but it is worth mentioning that the same approach can be applied to measurement tomography \cite{blumoffImplementingCharacterizingPrecise2016, chenDetectorTomographyIBM2019}, Hamiltonian tomography \cite{schirmerTwoqubitHamiltonianTomography2009, coleHamiltonianTomographyQuantum2015}, and gate-set tomography \cite{cincioMachineLearningNoiseResilient2021,nielsenGateSetTomography2021}.

There are promising approaches for QPT, including maximum-likelihood estimation \cite{fiurasekMaximumlikelihoodEstimationQuantum2001,sacchiMaximumlikelihoodReconstructionCompletely2001}, Bayesian estimation \cite{schultzExponentialFamiliesBayesian2019, lukensPracticalEfficientApproach2020}, projected gradient descent \cite{kneeQuantumProcessTomography2018}, projected least squares \cite{gutaFastStateTomography2020, surawy-stepneyProjectedLeastSquaresQuantum2022}, compressed sensing through convex optimization \cite{grossQuantumStateTomography2010, baldwinQuantumProcessTomography2014, rodionovCompressedSensingQuantum2014}, variational QPT \cite{torlaiQuantumProcessTomography2023}, and trace regression models \cite{kadriPartialTraceRegression2020}. 
The key challenge, however, is the exponentially growing size of the process representation, e.g., the Choi representation is a complex-valued $4^n \times 4^n$ matrix for $n$ qubits \cite{choiCompletelyPositiveLinear1975}. This makes it difficult to not only estimate the process from noisy data, but also to derive the source of noise and its interpretation. Furthermore, to ensure high-fidelity results, the informationally-complete set (ICS)
of measurements and input states are required, which also grows exponentially in size. As a result, all experiments so far have predominantly applied QPT to a maximum of 3 qubits \cite{AbuGhanem_2024}. Recently, methods such as ShadowQPT have emerged, which relax these constraints by approximating the quantum channel with classical shadows. Although these methods have successfully been performed on 4-qubit systems \cite{Levy_Luo_Clark_2024}, their solution lacks the properties of a quantum channel, making them unphysical.


To address the scalability concerns, we utilize a strategy similar to deep learning, which has proven to be successful in various areas of physics \cite{carleoSolvingQuantumManybody2017,carrasquillaReconstructingQuantumStates2019, sharirDeepAutoregressiveModels2020, ahmedGradientDescentQuantumProcess2023}. Deep learning involves optimizing complex computational models, such as neural networks, tensor networks \cite{liaoDifferentiableProgrammingTensor2019, panContractingArbitraryTensor2020}, and differential equation solvers \cite{torlaiWavefunctionPositivizationAutomatic2020}, using gradient-based methods. Automatic differentiation enables the computational process to be differentiable with respect to its constituent elements, such as a hidden layer in deep learning. As a result, the computational model can be fine-tuned with automatic differentiation using gradient-based techniques. 
In the context of quantum physics and QPT, the elements that are typically optimized are often limited to a specific class, such as the underlying group or symmetry structure.
Therefore, any optimization that is considered valid must comply with the corresponding constraints. This pertains to Riemannian optimization, which alters gradient-based methods to guarantee that the intermediate solutions maintain the desired property, such as unitarity, completely positive trace preserving (CPTP), etc. 

In particular, we model this process geometrically, as visualized in Fig.~\ref{fig:manifold}. All possible quantum channels exist on a smooth continuous surface known as a manifold $\mathcal{M}$. Gradient-based methods find the point that minimizes the loss function by taking steps along the manifold. In our case, the loss function compares simulated data generated by applying the current channel $\Bbbk$ with the real data that was generated by the true target channel $\mathcal{E}$. Furthermore, the unique properties of quantum channels restrict which channels are physically possible, changing the shape of our manifold. Specifically, quantum channels exist on the Steifel manifold.

This paper makes the following major contributions:
\begin{itemize}
    \item Formulates Riemmannian optimization via gradient descent on the Stiefel manifold for QPT.
    \item Adapts stochastic Adam optimizer for QPT.
    \item Provides an open-source implementation Qutee.jl \cite{volyaRustyBambooQuteeJl2023}, with native accelerated computation via GPUs.
    \item Demonstrates the effectiveness of QPT with respect to process dimension and characterization  on a set of real and simulated experiments.
\end{itemize}

The remainder of this paper is organized as follows. Section~\ref{sec:channel-representation} provides formulation of quantum channels. Section~\ref{sec:approach} describes our proposed QPT framework. Section~\ref{sec:simulation} presents two case studies using simulation of quantum processes. Section~\ref{sec:steady-state} presents a case study with hardware-based evaluation of our QPT framework. Finally, Section~\ref{sec:conclusion} concludes the paper.

\section{Quantum Channels} 

\label{sec:channel-representation}
A quantum channel is generally represented as a completely positive (CP) and trace-preserving (TP) linear map $\mathcal{E}$ that maps a quantum state represented by the density matrix $\rho$ to another state represented by $\rho^\prime$, i.e., $\rho^\prime = \mathcal{E}(\rho)$. The adjoint of a quantum channel $\mathcal{E}^\dagger$ generalizes the Heisenberg picture (under the Frobeniuus norm) and maps operators $E$ to other operators $E^\prime$, i.e., $E^\prime = \mathcal{E}^\dagger(E)$. This section details how $\mathcal{E}$ can be represented in various forms to highlight specific attributes.

\subsection{Kraus Representation}
For computations, it is convenient to use the Kraus-operator representation (also known as the canonical form) of a quantum channel $\mathcal{E}$ and its adjoint $\mathcal{E}^\dagger$. Formally, it is 
\begin{equation}\label{eq:kraus}
    \mathcal E(\rho) = \sum_{l = 1}^{k} {K_l} \rho {K_l}^{\dagger} \quad \text{ and } \quad  \mathcal E^\dagger(E) = \sum_{l = 1}^{k} {K_l}^\dagger E {K_l}
\end{equation}
where $\mathcal{E}$ acts on states $\rho$ and $\mathcal{E}^\dagger$ acts on operators $E$.
The $k$ complex-valued matrices \{$K_l$\} of dimension $N \times N$ are called the Kraus operators, with the requirement that the channel is trace-preserving (for $I$ identity) $\mathcal{E}^\dagger(I) = \sum_l K_l^\dagger K_l = I$.
Note that the Kraus representation is not unique, as $F_i = \sum_j U_{i,j}K_j$ represents the same CPTP map defined by a  unitary transformation $U$.

Additionally, it is more efficient to work with the minimum number of Kraus operators needed to represent a CPTP map, which is known as the \textit{Kraus rank}.
Given that a Hilbert space of dimension $N$ contains $N^2$ linearly independent operators, the Kraus rank is no larger than $N^2$ \cite{choiCompletelyPositiveLinear1975}.
Thus, we may convert the Kraus representation to an $N^2 \times N^2$ (Hermitian) Choi matrix and obtain the minimal representation by considering only eigenvectors with non-zero eigenvalues \cite{choiCompletelyPositiveLinear1975}. An alternative method is to compute the overlap matrix $C_{ij} = \mathrm{Tr}(K_iK_j^\dagger)$ and diagonalize it $C = V^\dagger D V$.
The new Kraus operators $\mathcal{\tilde{K}}_i = \sum_{j}V_{ij}K_j$ will be the minimal representation, where some matrices will be zeros if the original representation was redundant.
Finally, we define a matrix of dimension $kN\times N$
\begin{equation}\label{eq:kraus-stiefel}
    \Bbbk = [K_1\, K_2\, \hdots\, K_k]^T
\end{equation}
by column-stacking the $k$ Kraus operators.
This matrix $\Bbbk$ is an element of the Stiefel manifold, satisfying $\Bbbk^\dagger\Bbbk = I$ (the TP condition).

\begin{figure}[t]
    \centering
    \includegraphics{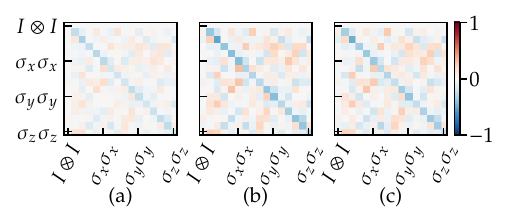}
    \vspace{-0.2in}
    \caption{Pauli-Liouville representation of simulated ``noise." A channel with Pauli noise $K_\mathrm{noise} = [\sqrt{p}I\otimes I, \sqrt{(1-p)}\sigma_z\otimes\sigma_z]$ is applied after a unitary operation $U$. Process tomography is performed to fit the noise+$U$ channel from measurement data. The ``noise" part defined as $\hat{P}\circ U^{-1} - \mathrm{I}$. (a) $p=0.1$, (b) $p=0.25$, and (c) $p=0.25$ with measurement data perturbed by a normal distribution with $\epsilon=0.01$.}
    \label{fig:pauliliou}
    \vspace{-0.1in}
\end{figure}

\subsection{Liouville and Choi Representation}
Although we perform all computations using the Kraus representation, it is illustrative to discuss the properties of CPTP maps via their Liouville representation. 
The Liouville (superoperator) representation of $\mathcal E$ is an $N^2\times N^2$ matrix of the form $\hat{\mathcal{E}} = \sum_l K_l \otimes \bar{K_l}$ ($\bar{K_l}$ being the complex conjugate of $K_l$) acting on a vectorized $N^2$-dimensional density matrix $| \rho \rangle\!\rangle$ from the left.
This representation, equivalent to the Kraus representation, conveniently allows for the examination of a channel's spectral properties along with performing channel composition, i.e., for channels $\mathcal{E}$ and $\mathcal{L}$: $(\mathcal{E}\circ\mathcal{L})(\rho) = \mathcal{E}(\mathcal{L}(\rho)) = \mathcal{\hat{E}}\cdot\mathcal{\hat{L}}\cdot| \rho \rangle\!\rangle$ where $\cdot$ is the standard matrix (vector) product. As a result, a specific component of the quantum process, such as noise, can be isolated.

Since the Liouville representation consists of complex elements, it can be convenient to represent $\mathcal{\hat{E}}$ (when $N = 4^n$) in the Pauli basis, where all the matrix elements are real. The Pauli-Liouville may be written as $\mathcal{\hat{P}}_{i,j} = \mathrm{Tr}\left[E_j^\dagger\mathcal{E}(E_i)\right]$ where operators $\{E_i\}$ are the Pauli operators. 
For example, Fig.~\ref{fig:pauliliou} displays a heatmap of the Pauli-Liouville matrix where a unitary channel is extracted to reveal only a ``noise" part. 

Furthermore, the Liouville matrix is related to the Choi matrix $\hat{J}$ through an involution $\hat{J}_{ij,kl} = \mathcal{\hat{E}}_{ik,jl}$.
By the Choi-Jamiołkowski isomorphism, a quantum channel may be viewed as a quantum state. Therefore, one way to define distances between channels is via state fidelity
\vspace{-0.05in}
\begin{equation}\label{eq:fidelity}
    F(\rho,\sigma) = \frac{1}{N^2}\left(\mathrm{Tr}\sqrt{\sqrt{\rho}\sigma\sqrt{\rho}}\right)^2
    \vspace{-0.05in}
\end{equation}
where $\rho$ and $\sigma$ are two quantum channels as Choi matrices.

\section{Quantum Process Tomography using Riemmanian Optimization}
\label{sec:approach}

This section describes our proposed QPT framework via Riemmanian optimization.
Assuming the convention in (\ref{eq:kraus-stiefel}) of column-stacked Kraus operators that reside in the Stiefel manifold $\mathcal{M}$, and a loss function $\ell(\cdot)$, we formulate the task of QPT as the following optimization problem
\begin{equation}\label{eq:opt}
\min_{\Bbbk \in \mathcal{M}} \ell(\Bbbk).
\end{equation}
A general loss function for the QPT task may be defined as
\vspace{-0.05in}
\begin{equation}\label{eq:loss}
    \ell(\Bbbk) = \sum_{i,j} \left[d_{i,j} - \mathrm{Tr}\left(M_j\sum_c K_c \rho_i K_c^\dagger\right)\right]^2,
    \vspace{-0.05in}
\end{equation}
where $M_j$ denotes the $j$-th measurement operator, $\rho_i$ denotes the $i$-th initial state, and $d_{i,j}$ denotes readout of the experiment.
In other words, the optimization problem is to find a set of Kraus operators that minimizes the difference between experimental and simulated data.
A graphical representation of this process is presented in Fig.~\ref{fig:gradDescent}.

This section is organized as follows. First, we outline the steps for using the output of automatic differentiation to obtain a gradient residing in the tangent bundle.
Next, we discuss an efficient method for performing the gradient descent via retraction.
Finally, we discuss optimization opportunities. 

\subsection{Stiefel Manifold, Tangents, and Automatic Differentiation}

We recall that the Stiefel manifold consists of all $n\times p,\,n\geq p$ unitary matrices 
\begin{equation}
    \mathcal{M}(n,p) = \{\Bbbk \in \mathbb{C}^{n\times p}: \Bbbk^\dagger \Bbbk = I_p\}.
\end{equation}
The tangent space at a point $\Bbbk \in \mathcal{M}(n,p)$ is given by
\begin{equation}\label{eq:tangent-space}
    T_\Bbbk \mathcal{M}(n,p) = \{ \Bbbk\Omega + \Bbbk_{\perp}\Theta: \Omega = -\Omega^\dagger, \Theta \in \mathbb{C}^{(n-p)\times p} \}
\end{equation}
where $\Bbbk_\perp \in \mathbb{C}^{n\times (n-p)}$ is orthonormal and $\mathrm{span}(\Bbbk_\perp) = (\mathrm{span}(\Bbbk))^\perp$.
The Stiefel manifold becomes a Riemannian manifold by introducing a Riemannian metric $g$, which is a smoothly varying inner product in the tangent space.
There are two natural inner products for the tangent space: the Euclidean inner product inherited from the embedding space $\mathbb{C}^{n\times p}$ and the canonical inner product. That is, given $A,B \in T_\Bbbk \mathcal{M}(n,p)$, we may opt for either:
\begin{equation}
        \langle A,B \rangle = \left\{\mathrm{Tr}\left(A^\dagger B\right) \,\, \mathrm{or} \,\,  \mathrm{Tr}\left(A^\dagger (I - \frac{1}{2}XX^\dagger)B\right) \right\}.
\end{equation}
The distinction arises in the norm of a vector $A = \Bbbk\Omega + \Bbbk_{\perp}\Theta$,
\begin{equation}
    ||A|| = \sqrt{\langle A,A \rangle} = \begin{cases}
    \sqrt{||\Omega||^2 + ||\Theta||^2}\,\,\,\,\, \mathrm{(inherited)} \\
    \sqrt{\frac{1}{2}||\Omega||^2 + ||\Theta||^2} \, \mathrm{(canonical)}
\end{cases}
\end{equation}
The factor of $\frac{1}{2}$ gives an equal weighting of both $\Omega$ and $\Theta$ ($\Omega = -\Omega^\dagger$ produces an extra factor of $2$.) We assume the canonical inner product throughout the remainder of this paper.

We use reverse (adjoint) mode automatic differentiation to compute the loss gradient $\ell: \mathcal{M}(n,p) \to \mathbb{R}$, which we will denote by $\nabla\ell$.
However, automatic differentiation assumes that the underlying space is Euclidean.
Thus, the calculated gradient $\nabla\ell \in \mathbb{C}^{n\times p}$ does not adhere to the properties of tangent space of the Stiefel manifold in (\ref{eq:tangent-space}). 
To obtain a gradient that resides in the tangent space, we use the fact that the Stiefel manifold may be embedded in the Euclidean space.
Therefore, the gradient vector in the Stiefel tangent space can be expressed as
\vspace{-0.1in}
\begin{equation}
    \nabla_{T_\Bbbk\mathcal{M}} \ell = \mathrm{proj}_\Bbbk(\nabla\ell)
\end{equation}
where $\mathrm{proj}_\Bbbk$ projects the Euclidean gradient onto the Stiefel tangent space and is defined as
\begin{equation}\label{eq:projection}
    \mathrm{proj}_\Bbbk(X) = X - \Bbbk (\Bbbk X^\dagger + \Bbbk^\dagger X)/2.
\end{equation}

\subsection{Gradient Descent and Retraction}

Gradient-based methods typically assume an initial guess $x$ for a parameter and then iteratively evaluate the loss function, compute the gradient, and then update the parameter
\begin{equation}\label{eq:gradient}
    x \leftarrow x - \beta \nabla \ell(x).
\end{equation}
Similarly, the gradient-based approach to solve the optimization problem in (\ref{eq:opt})  assumes an initial guess for the quantum channel $\Bbbk \in \mathcal{M}$ and iteratively updates the guess.
However, unlike the simple update rule of (\ref{eq:gradient}), extra care must be taken to ensure that each individual update to $\Bbbk$ remains on the manifold.

Let $\gamma(\tau)$ denote a curve on the manifold such that $\gamma(0)=\Bbbk$ and $\dot\gamma(0) = \nabla_{T_\Bbbk\mathcal{M}}\ell(\Bbbk)$.
The curve $\gamma(\tau)$ is considered a geodesic if it satisfies the exponential map
\begin{equation}
    \gamma(\tau) = \mathrm{Exp}_\Bbbk(\tau \nabla_{T_\Bbbk\mathcal{M}}\ell(\Bbbk)),\, \mathrm{for}\, \tau\in[0,1].
\end{equation}
In other words, the exponential map takes a vector in tangent space to a point on the manifold such that it remains on the shortest path.
Recall that the objective is to minimize a function while remaining constrained to the manifold.
One way to view this is to find the geodesic on the manifold that moves us from the starting point to the optimal point that minimizes the function.
However, this is difficult to formulate generally, and instead we must rely on approximate algorithms such as line search.
In this case, via the gradient-based method, we perform a step in the tangent space and are still constrained to the manifold via the exponential mapping.

The exponential map is not computationally efficient.
We can approximate the exponential map to $m$-th order in a Taylor or Pad\'e expansion.
Other approximations include matrix decomposition, such as QR and singular value decomposition (SVD).
Such approximation of the exponential is known as retraction.
A common choice for a retraction is the Cayley transform (a first-order Pad\'e expansion), which defines a curve
\begin{equation}\label{eq:cayley}
    Y(\tau) = \left(I + \frac{\tau}{2} W\right)^{-1}\left(I - \frac{\tau}{2} W\right) \Bbbk
\end{equation}
where { $W$} is a skew-hermitian matrix, i.e. { $W^\dagger = -W$}. By choosing { $W = \bar{W} - \bar{W}^\dagger$}, where { $\bar{W} = \nabla \ell(\Bbbk) \Bbbk^\dagger - \frac{1}{2} \Bbbk (\Bbbk^\dagger \nabla \ell(\Bbbk) \Bbbk^\dagger)$}, the transformation implicitly projects the gradient onto the tangent space in the descent direction. Here, { $W$} is the matrix operator for projection in (\ref{eq:projection}).

\begin{figure}
    \centering
        \centering
        \includegraphics[width=0.7\linewidth]{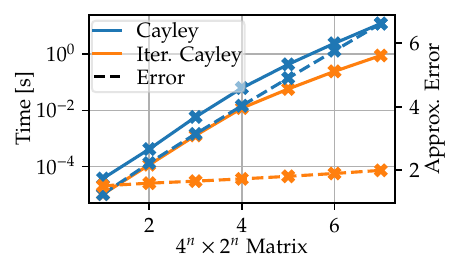}
        \vspace{-0.1in}
        \caption{ Iterative-based Cayley transform has better performance (time and accuracy) with respect to traditional Cayley transform. The accuracy (dashed lines) are computed by solving the exact exponential map and taking the norm difference with respect to approximated retraction. }
        \vspace{-0.2in}
    \label{fig:time}
\end{figure}

For large models $\Bbbk$, the Cayley transform in (\ref{eq:cayley}) cannot be performed efficiently due to the matrix inversion.
It is possible to simplify the computation of the inverse via the Sherman-Morrison-Woodbury formula, however, this is only efficient to do so for matrices where $n \ll p$.
Instead, we adapt the fixed-point iteration method for the Cayley transform \cite{liEfficientRiemannianOptimization2020}, which requires only matrix multiplications while also serving as an efficient approximation.
The iterative update to the Cayley transform is defined as
\begin{equation}\label{eq:iter-cayley}
    Y(\tau)^i \leftarrow \Bbbk + \frac{\tau}{2}W(\Bbbk + Y(\tau)^{i-1})
\end{equation}
where $i$ denotes the iterative step.
Figure~\ref{fig:time} highlights the computational and error improvement in iterative Cayley.

\subsection{Momentum Updates and Vector Transport}

Data collected in the real world are subject to randomness. This means that the measurements we obtain are assumed to come from an underlying probability distribution. Therefore, our objective is to minimize the expected value $\mathbb{E}[\ell(\Bbbk)]$, where $\ell(\cdot)$ depends on stochastic measurement data $d_{i,j}$. We may consider the gradient $\nabla \ell$ as a random variable for which obtaining the true first (mean) and second (standard deviation) moments is costly. Adam, a successful technique for gradient-based optimization of stochastic objective functions, estimates and updates the moments via random subsamples (minibatches) of data points. Adam further accumulates the gradients from previous iterations into a momentum term, which reduces the oscillations in $\ell(\cdot)$ arising from randomness and hence accelerates convergence. However, Adam is only suitable for optimizations in a Euclidean space.  We adopt Adam for the task of QPT on the Stiefel manifold. 

The key idea is to move a vector in the tangent space along the search paths on the manifold. In general, this is done using a parallel transport that preserves the geometrical properties of the vector from one tangent space to another. However, this requires the solution of a costly differential equation. Instead, some geometric conditions may be relaxed to perform an efficient \textit{vector transport}. Since the Stiefel manifold is a submanifold of the Euclidean space, the tangent space is a subspace of the Euclidean space. Hence, a vector transport on the Stiefel manifold is a projection onto the tangent space. Following \cite{liEfficientRiemannianOptimization2020}, using the property that the projection is a linear map, the key Adam steps may first be performed in Euclidean space, followed by a projection:
\begin{equation}
\nonumber
    \alpha \mathrm{proj}_{\Bbbk_{k}}(M_k) + \beta \mathrm{proj}_{\Bbbk_{k}}(\nabla \ell(\Bbbk_k)) = \mathrm{proj}_{\Bbbk_{k}}(\alpha M_k + \beta \nabla\ell(\Bbbk_k))
\end{equation}
Namely, a linear combination of the gradient $\nabla \ell(\Bbbk)$ and the momentum $M_k$ for step $k-1$ is taken with real ``learning" coefficients $\alpha$ and $\beta$. Instead of explicitly performing a projection, the iterative Cayley transform in (\ref{eq:iter-cayley}) implicitly projects the vector onto the tangent space while also performing a parameter update.

\section{Case Study: Simulation of Quantum Processes}
\label{sec:simulation}

In this section, we present the results of our QPT optimization on a set of problems with varying difficulty: randomized channels and controlling coherent states of a harmonic oscillator. In both  examples, we use a random initial guess $\Bbbk$ for the channel.
Furthermore, all optimizations are performed on an Nvidia RTX6000 GPU with 48GB of GDDR6 of memory. 





\subsection{Random Channels with Insufficient Measurements}

In our first case study, we explore the capabilities of Riemannian optimization of quantum processes with respect to dimensions, Kraus rank, and measurement error.
We follow a direct QPT approach, where $6^n$ probes $\rho_i$ and $6^n$ measurements $M_j$ are selected as eigenstates of the generalized Gellmann matrices.
For a random channel $\mathcal{E}_\mathrm{rand}$, the simulated data is computed as
\begin{equation}
    d_{i,j} = \mathrm{Tr}(M_j \mathcal{E}_{\mathrm{rand}}(\rho_i)) + \mathcal{N}(0, \epsilon)
\end{equation}
where $\mathcal{N}(0,\epsilon)$ introduces measurement error following a normal distribution with variance $\epsilon$.
We generate a random channel $\Bbbk$ by computing the orthonormal part of the QR decomposition of a random $rn\times n$ matrix where each element is sampled from the normal distribution.  

\begin{figure}[t]
    \centering
    \begin{subfigure}[b]{0.49\linewidth}
        \centering
        \includegraphics[width=\linewidth]{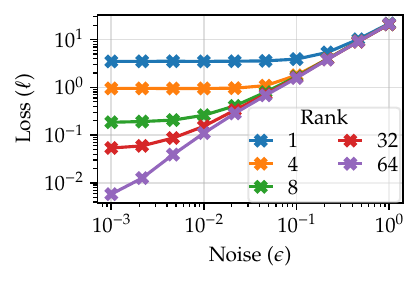}
        \vspace{-0.2in}
        \caption{Mean Loss}
        \label{fig:error-versus-k}
    \end{subfigure}
    \hfill
    \begin{subfigure}[b]{0.49\linewidth}
        \centering
        \includegraphics[width=\linewidth]{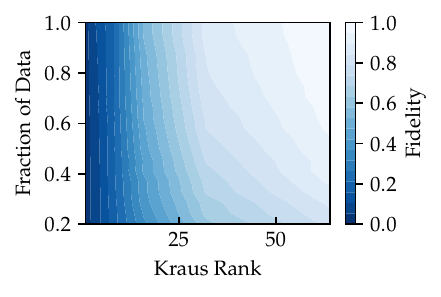}
        \vspace{-0.2in}
        \caption{Channel Fidelity}
        \label{fig:data-versus-k}
    \end{subfigure}
    \caption{(a) Mean loss of the optimization as noise $\epsilon$ is increased for different channel ranks. Reconstruction error saturates as a power curve with respect to noise $\epsilon$ (note the log-log scale.) (b) Channel fidelity at $\epsilon=0.01$ for a reconstructed quantum channel with varying Kraus rank as well as access to only a fraction of data. Kraus rank is the dominant factor for fidelity, while the fraction of data is responsible for fidelity margin.}
\end{figure}

Figure~\ref{fig:error-versus-k} shows simulation results for a 3-qubit system undergoing a random process $\mathcal{E}_\mathrm{rand}$.
The Hilbert space is $2^3=8$, with a maximal Kraus rank of $4^3=64$.
Various Kraus ranks are fitted, with different measurement noise.
The computed loss with lower Kraus ranks saturate sooner, as the fitted channel is simply not expressive enough to represent the full-rank process.
Figure~\ref{fig:data-versus-k} further exemplifies the importance of Kraus rank. The fidelity of the reconstructed channel (\ref{eq:fidelity}) with respect to original channel is computed, where the reconstructed channel was computed with a fraction of data, namely $\sqrt{\nu} 6^n$ probes and $\sqrt{\nu}6^n$ measurements are used given a ratio $\nu\in(0,1]$. High fidelity is observed even with the absence of a fraction of data.
We are able to reconstruct the quantum channel in a few seconds, even at maximal rank.
In comparison, other tomography techniques become limiting even at 3-qubits. Table \ref{tab:qpt_time} compares our method to state-of-the-art libraries, such as Qiskit \cite{Qiskit} (via constrained maximum-likelihood estimate), for several qubit sizes. 

\begin{table}[t]
    \setlength{\tabcolsep}{3pt}
    \centering
        \caption{Comparing execution time of QPT methods.}
    \begin{tabular}{||c c c c c||}
        \hline
        \# Qubits & Hilbert space & Basis size & \textbf{Qiskit time} & \textbf{Our time} \\
        \hline\hline
        2 & 4 & 16 & 4.153s & 3.424s \\
        3 & 8 & 64 & 76.134s & 4.791s \\ 
        4 & 16 & 256 & 1355.841s & 61.766s \\
        \hline
    \end{tabular}
    \label{tab:qpt_time}
    \vspace{-0.1in}
\end{table}

\subsection{Universal Control of a Harmonic Oscillator}\label{sec:oscilattor}

Our second case study considers a continuous-variable quantum system, where information is encoded in an infinite-dimensional Hilbert space, following a similar example in~\cite{ahmedGradientDescentQuantumProcess2023}. Due to the nature of an infinite-dimensional system, selecting an appropriate Hilbert space cut-off is crucial for correctly describing a state numerically.
In our case, we consider a Fock space cut-off of $N=64$, which exceeds previously considered dimension cut-offs.

\begin{figure}[t]
    \centering
    \includegraphics[width=\linewidth]{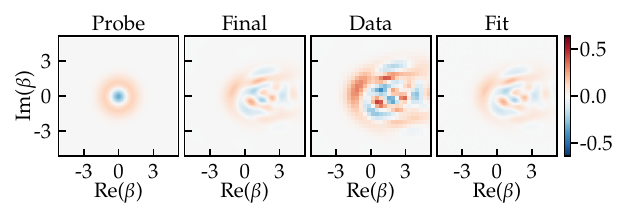}
    \vspace{-0.2in}
    \caption{Wigner functions of a probe state undergoing ideal simulation and tomographic reconstruction with a Fock space $N=64$. The data are measured values of displaced parity operator $\Pi_d(\beta)$, but with fewer sample points $\beta$ to mimic an experimental setup.}
    \label{fig:wigner}
    \vspace{-0.1in}
\end{figure}

Such systems (single bosonic mode) may be modeled as quantum harmonic oscillators with associated creation $\hat{\alpha}^\dagger$ and annihilation operators $\hat{\alpha}$.
Coherent states $\ket{\alpha}$ are the unique eigenstates with corresponding eigenvalue $\alpha$ of the annilation operator $\hat{\alpha}\ket{\alpha} = \alpha\ket{\alpha}$.
Universal control of a harmonic oscillator can be achieved by combining displacement operation
\begin{equation}
    D(\alpha) = \exp(\alpha \hat{\alpha}^\dagger - \alpha^*\hat{\alpha}),\quad \ket{\alpha}=D(\alpha)\ket{0}
\end{equation}
and a selective number-dependent arbitrary phase (SNAP) gate 
\begin{equation}
    S(\vec{\phi}) = \sum_n e^{i\phi_n}\ket{n}\bra{n} 
\end{equation}
which adds a real phase $\phi_n$ to the Fock state $\ket{n}$.
This can be shown by observing that the generators and commutators of $D(x)$ and $S(\phi)$ generate the full Lie algebra $u(N)$ for any truncated oscillator space \cite{Krastanov_2015}.

We simulate a setup where the coherent probe states $\rho_i = \ket{\alpha}\bra{\alpha}$ undergo a displacement + SNAP process, and are then measured via a displaced-parity operator \cite{Bishop_1994}
\begin{equation}
    \Pi_d(\beta) =\sum_{n=0}^N (-1)^i  D(\beta)\ket{n}\bra{n} D(\beta)^\dagger.
\end{equation}
Similarly, the Wigner functions are approximated using expectation values of the parity operator \cite{Royer_1977}.
Figure~\ref{fig:wigner} shows the Wigner functions for one instance of the complete process.
In other words, an initial probe state $\rho_i$ undergoes the simulated process to a final state $\rho_i^\prime$.
The computation of the expectation values for the displaced-parity operator is performed on a discretized grid.
This was done to mimic an experimental scenario in which conducting a full Wigner tomography is infeasible.
We find good results with a rank-fitted channel $r=2$, despite the fact that the process is purely unitary (r=1).
This indicates, non-unitary behavior, which arises from the discretization of measurement expectation values.

\section{Steady States of Nondemolishment Measurement in Quantum Hardware}
\label{sec:steady-state}

\begin{figure}[t]
    \centering
    \input{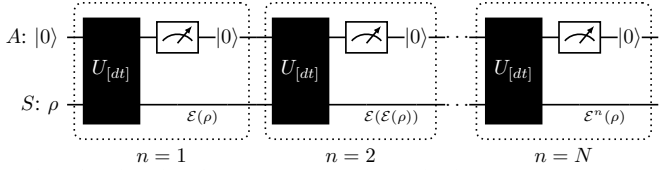}
    \vspace{-0.1in}
    \caption{Circuit of quantum nondemolishing measurement.}
    \vspace{-0.1in}
    \label{fig:blackbox}
\end{figure}

In the following experiment, we assume a setup as shown in Fig.~\ref{fig:blackbox}.
Namely, a system that is coupled to a detector
\begin{equation}
    \rho = \ket{0}_D\bra{0}_D\otimes\rho_S
\end{equation}
where the system is in an unknown mixed state $\rho_S$ and the detector is in a known pure state $\ket{0}_D$.
The system evolves for a duration $dt$ subject to a two-body Hamiltonian $H$,
\begin{equation}
\rho^\prime = U \rho U^\dagger,\quad U = \exp(-iHdt).    
\end{equation}
The detector is then measured and promptly reset to its known state $\ket{0}$.
As a result of this procedure, the system is disrupted, resulting in a new state given by a partial trace
\begin{equation}
    \rho_S = \mathrm{Tr}_{D}\left[U (\ket{0}_D\bra{0}_D\otimes\rho_S) U^\dagger\right].
\end{equation}
In this ideal scenario, we may express the evolution of $\rho_S$ as a quantum channel $\mathcal{E}$ in terms of two Kraus operators
\begin{equation}\label{eq:kraus-ideal}
    K_1 = \bra{0}_D U \ket{0}_S \text{ and } K_2 = \bra{0}_D U \ket{1}_S.
\end{equation}
The evolution then follows (\ref{eq:kraus}). The repeated application of $\mathcal{E}$, will cause the system to converge to a steady state.

We run the circuit in Fig.~\ref{fig:blackbox} on three different IBM Quantum computers: \emph{ibmq\_lima}, \emph{ibmq\_belem}, \emph{ibm\_perth}.
The system and detector are selected as neighboring qubits coupled via a cross-resonance (CR) gate.
We idle the qubits for a duration $dt$, corresponding to an idealized evolution $U_{[dt]}$.
After the detector measurement, the system is measured on a different basis by performing a single qubit rotation gate followed by measurement in the computational basis.
This process is repeated in which the measurement of the system is only performed in the end, providing an estimate of the quantum state $\rho_S$ after each non-demolishing measurement.
In addition, we performed 30 trials.

\begin{figure}[t]
    \centering
     \begin{subfigure}[b]{\linewidth}
         \centering
         \includegraphics[width=\textwidth]{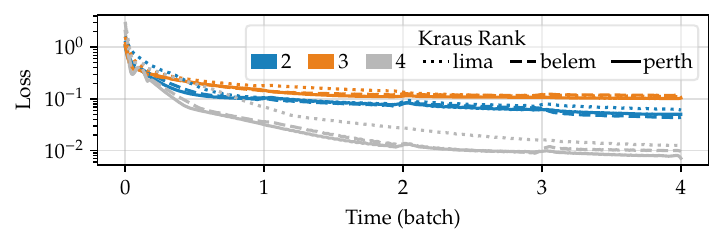}
         \caption{History of the loss $\ell(\cdot)$ throughout optimization with varying Kraus rank.}
     \end{subfigure}
     \begin{subfigure}[b]{\linewidth}
        \centering
    \includegraphics[width=1\linewidth]{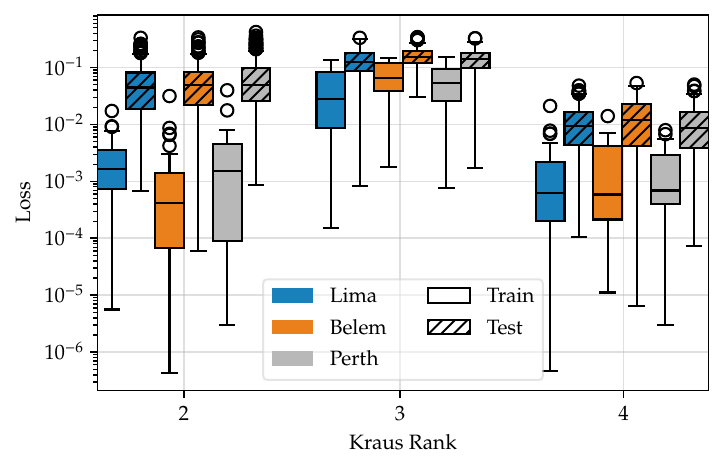}
    \caption{Box-and-whisker plots of the final loss with varying Kraus ranks for training and testing sets of data.}
    \label{fig:rank}
    \end{subfigure}
    \caption{Results of optimization on IBM Quantum computers.}
    \vspace{-0.1in}
\end{figure}

The process is considered to be strictly Markovian, and that $U_{[dt]}$ is stable from one iteration to the next.
Assuming the loss function of (\ref{eq:loss}), we fit a channel $\mathcal{E}$ with ranks $r=\{2,3,4\}$ in 20 trials, leaving 10 trials to test the fitted channel.
As shown in Fig.~\ref{fig:rank}, the maximum rank ($r=4$) provides the best results from optimization against the test data.
While (\ref{eq:kraus-ideal}) states that the ideal scenario is described with a Kraus rank $r=2$, in reality the detector state is not perfectly initialized to $\ket{0}$ and may instead be in a state $\ket{1}$ or a general superposition. Hence, the maximum rank of $r=4$ is required to capture faulty initialization.
In addition, stochastic and non-Markovian effects are present on quantum devices which can no longer be described by a quantum channel.

\section{Conclusion and Future Work}
\label{sec:conclusion}

We have introduced a data-driven approach for the constrained optimization task of quantum process tomography.
Utilizing advanced stochastic objective optimizers, we can achieve outstanding results by considering the quantum process as residing in a Stiefel manifold and performing local updates that respect the geometry.
We implement an iterative Cayley transform for an efficient retraction on the manifold, as well as adopt the implicit projection to include moment estimates of stochastic optimization.
These considerations alleviate some numerical and memory issues, which distinguishes this approach from other tomographic methods. 
We demonstrated that our approach can quickly reconstruct quantum processes, such as scenarios with a Hilbert space of dimension 64.
Furthermore, using random processes, we established the relationship between Kraus rank and measurement error for precision in estimation.

There are numerous promising avenues for future research.
One interesting avenue is applying methods of Riemmanian optimization to other approximations and descriptions of quantum processes, such as tensor networks.
Namely, such avenues may be utilized to characterize and deduce non-Markovian noise processes.
Another important avenue is the consideration of error bars and uncertainty.
In part, such considerations are important for distinguishing errors and confidence regions of the tomographic process itself.   

\section*{Acknowledgments}
We acknowledge the use of IBM Quantum services for this work. The views expressed are those of the authors, and do not reflect the official policy or position of IBM or the IBM Quantum team.
The computing systems used in this work is supported by DOE (DE-SC-0009883).  
This research is partially supported by the grants from DARPA (HR0011-24-3-0004) and NSF (CCF-1908131).

\balance
\bibliographystyle{IEEEtran}
\bibliography{references}

\end{document}